\documentclass[final,5p,times]{elsarticle}

\usepackage{amsmath,amsfonts,latexsym,amssymb,graphicx,makeidx,cases}
\usepackage{xcolor,diagbox,enumitem}
\usepackage{multirow}
\usepackage{cuted}
\usepackage{capt-of}
\usepackage[colorlinks,linkcolor=blue,anchorcolor=blue,citecolor=blue,urlcolor=blue]{hyperref}
\usepackage{mathrsfs}
\usepackage{orcidlink}
\biboptions{sort&compress}
\journal{Science China Physics, Mechanics \& Astronomy}

\begin{document}

\begin{frontmatter}

\title{Spectral Butterfly Effect and Resilient Ringdown in Thick Braneworlds}

\author[aff1,aff2]{Hai-Long Jia\orcidlink{0009-0009-2479-4005}}
\author[aff1,aff2]{Wen-Di Guo\orcidlink{0000-0003-2145-6587}}
\author[aff1,aff2]{Yu-Peng Zhang\orcidlink{0009-0008-3632-8194}}
\author[aff1,aff2]{Yu-Xiao Liu\corref{cor1}\orcidlink{0000-0002-4117-4176}}
\ead{liuyx@lzu.edu.cn}
\cortext[cor1]{Corresponding author. E-mail: liuyx@lzu.edu.cn.}

\affiliation[aff1]{organization={Lanzhou Center for Theoretical Physics, Key Laboratory of Theoretical Physics of Gansu Province, Key Laboratory of Quantum Theory and Applications of Ministry of Education, Gansu Provincial Research Center for Basic Disciplines of Quantum Physics, Lanzhou University},
            city={Lanzhou},
            postcode={730000},
            country={China}}
\affiliation[aff2]{organization={Institute of Theoretical Physics and Research Center of Gravitation, School of Physical Science and Technology, Lanzhou University},
            city={Lanzhou},
            postcode={730000},
            country={China}}

\begin{abstract}

    The quasinormal mode spectrum is a unique fingerprint 
    linking gravitational-wave observations to extra-dimensional geometry. 
    In this article, we show that thick braneworlds exhibit a spectral butterfly effect: 
    infinitesimal deformations of the effective potential trigger dramatic migrations of quasinormal modes, 
    challenging the presumed stability of this fingerprint. Frequency-domain instabilities depend 
    sensitively on the perturbation's location and strength. 
    In the time domain, near-brane perturbations primarily modify the early ringdown, 
    while far-brane perturbations generate clean late-time echoes. Crucially, 
    the graviton zero mode remains localized, preserving four-dimensional gravity. 
    Despite this pronounced spectral fragility, the observable early-stage signal 
    under current detector sensitivities is still dominated by the original fundamental mode. 
    Hence, thick braneworlds display a nontrivial coexistence of a fragile spectrum and a resilient ringdown, 
    supporting the continued use of the standard fingerprint in present-day 
    gravitational-wave astronomy while revealing its hidden sensitivity.

\end{abstract}


\end{frontmatter}

\allowdisplaybreaks

\section{Introduction}
The braneworld paradigm---exemplified by the Arkani-Hamed--Dimopoulos--Dvali (ADD)~\cite{Arkani-Hamed:1998jmv,Antoniadis:1998ig} 
and Randall--Sundrum~\cite{Randall:1999ee,Randall:1999vf} models---resolves the gauge hierarchy problem 
through a geometric mechanism rooted in higher-dimensional spacetime. 
Building on this concept, thick braneworlds provide a natural framework for localizing gravity 
on a codimension-one surface while embedding a smooth extra dimension~\cite{DeWolfe:1999cp,Gremm:1999pj,Csaki:2000fc,Herrera-Aguilar:2010ehj}. 
A key diagnostic of these models is the spectrum of tensor perturbations: 
the massless graviton zero mode recovers four-dimensional gravity, 
while the massive Kaluza--Klein (KK) sector---including quasinormal modes (QNMs) and long-lived resonances---encodes the bulk geometry~\cite{Seahra:2005wk,Seahra:2005iq,
Chung:2015mna,Tan:2023cra,Tan:2022vfe,Jia:2024pdk,Jia:2024sdk,Csaki:2000pp,Brevik:2002yj,Liu:2009ve,Zhong:2016iko,Zhu:2024gvl}. 
This KK fingerprint may be probed through gravitational-wave ringdown~\cite{Konoplya:2023fmh,NANOGrav:2023gor,Koyama:2004cf,Caprini:2018mtu}, 
collider signatures~\cite{Nath:1999mw,Bhattacharyya:2009br,Savina:2015zda,Agashe:2020wph}, 
and corrections to Newtonian gravity~\cite{Callin:2004py,Guo:2010az,Araujo:2011fm}.

The utility of this fingerprint rests on an implicit assumption of spectral stability: 
small deformations of the background should produce proportionally small shifts in the QNM spectrum. 
However, recent studies of black-hole spectroscopy have revealed a dramatic counterexample---the pseudospectrum instability---in which 
infinitesimal modifications of the effective potential can trigger large migrations of QNM frequencies~\cite{Jaramillo:2020tuu,Cheung:2021bol,Jaramillo:2021tmt,Berti:2022xfj}. 
This behavior has also been examined for quantum-corrected Schwarzschild and Boulware-Deser-Wheeler black holes~\cite{Cao:2024qcs,Cao:2025bdw}. 
Whether extra-dimensional spectral fingerprints exhibit a similar fragility remains largely unexplored. 
In this article, we ask whether representative smooth deformations that preserve localization already suffice to induce order-one rearrangements of the braneworld QNM spectrum.

The use of a thick brane is a deliberate choice rather than an assumption that other extra-dimensional scenarios are irrelevant. 
In ADD models with compact extra dimensions, the KK spectrum is controlled mainly by global compactification data and does not naturally provide a smooth volcano potential with a localized four-dimensional graviton on a noncompact extra dimension. 
Thin Randall--Sundrum models do provide gravitational localization, but the idealized brane is distributional and the perturbation problem is controlled by junction conditions at the brane. 
By contrast, a thick brane generated by bulk matter gives a smooth effective potential, a normalizable zero mode, and a massive QNM sector in which localized deformations of the geometry can be introduced without singular boundary terms. 
This makes it a clean testing ground for the question addressed here. 
More generally, in an Einstein--matter thick brane background, matter or geometric perturbations are encoded in the warp factor and therefore in the tensor superpotential $W$ and the tensor effective potential. 
By analyzing a canonical thick brane, we identify two broad effective-potential responses: near-brane reshaping and far-brane remote-structure formation. 
The parameterized Type I and Type II profiles used below are designed to represent these two responses, 
so the method and qualitative conclusions are expected to 
apply to localization-preserving Einstein--matter thick branes 
that exhibit these two classes of effective-potential responses. 
For thick brane models in modified gravity or other non-Einstein--matter frameworks, existing QNM calculations show that model-specific dynamics can alter the effective potential and spectrum~\cite{Li:2026ft}. 
The qualitative conclusions established here for localization-preserving Einstein--matter thick branes should therefore not be assumed to hold in such models without a separate, model-specific calculation; nevertheless, the present deformation-and-tracking strategy can be extended to analyze them.

We address this question for a canonical thick brane using two representative localized deformations of the volcano potential. 
These deformations provide a controlled way to encode smooth background reshaping while preserving the localization structure, and they capture the two physically distinct regimes most relevant for QNM migration.
Already in this controlled setting, the QNM frequencies undergo parametrically large migrations whose character depends strongly on locality. 
Near-brane deformations primarily destabilize higher overtones, while far-brane deformations can drive spiral migration, 
branch switching, and overtaking of the least-damped mode: the very mode that controls the longest-lived part of the ringdown response.

The time-domain lesson is subtler. 
With diagnostic wave packets chosen to isolate the QNM sector, the prompt waveform remains much less sensitive than the full spectrum. 
The graviton zero mode stays localized, preserving four-dimensional gravity. 
Near-brane perturbations produce controlled early-time shifts, whereas far-brane perturbations reveal themselves most clearly through delayed echoes rather than through a wholesale reorganization of the prompt signal.

Thick braneworlds thus exhibit a nontrivial coexistence of a fragile QNM spectrum and a resilient ringdown. 
The main lesson is therefore sharper than a simple stability/instability dichotomy: frequency-domain fragility need not imply equally strong finite-time waveform fragility, even though the hidden spectral sensitivity remains physically meaningful.

\section{Materials and methods}
We use the canonical thick brane generated by a single bulk scalar field~\cite{DeWolfe:1999cp,Gremm:1999pj}. 
After transforming to the conformal coordinate $z$, the extra-dimensional profile of the tensor perturbation obeys the Schr\"odinger-like equation
\begin{equation}
    \label{Slike-equation-same}
    \left[-\partial_{z}^{2} + V_0(z)\right] \psi(z) = m^2 \psi(z), 
\end{equation}
with the unperturbed volcano potential
\begin{equation}
    V_0(z)= \frac{3k^2\left(5k^2z^2 - 2\right)}{4 \left(k^2z^2 +1\right)^2}.
\end{equation}
Here $k$ sets the characteristic energy scale of the five-dimensional bulk.
This potential supports a normalizable graviton zero mode and a tower of massive KK excitations whose QNM spectrum serves as the extra-dimensional fingerprint.

\begin{figure*}[!t]
    \centering
    \includegraphics[width=\textwidth]{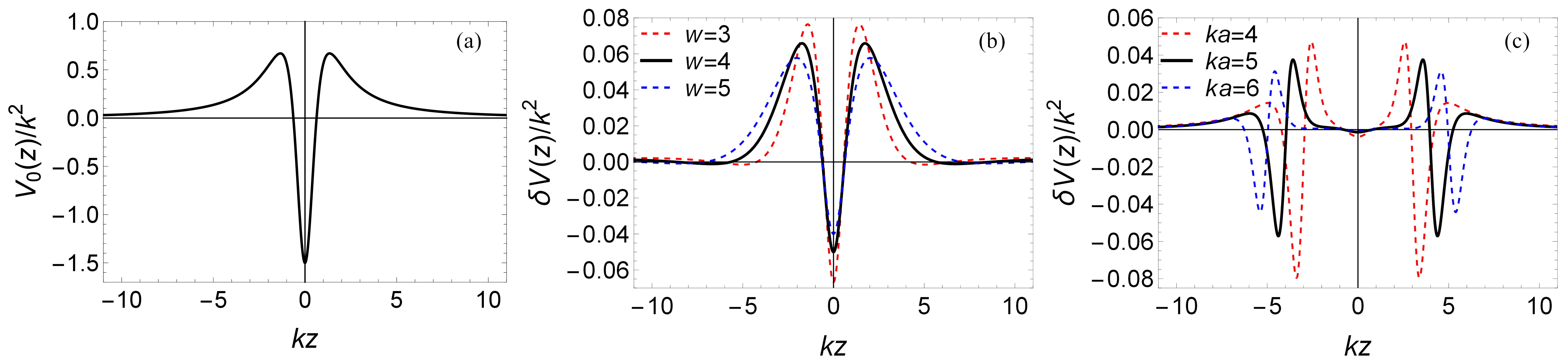}
    \caption{The unperturbed effective potential and two representative deformations: 
    (a) the unperturbed potential, (b) Type I perturbations, and (c) Type II perturbations. 
    The deformations are shown for $\epsilon=0.1$.}
    \label{figure-1}
\end{figure*}

A key observation is that $V_0(z)$ is not an independent input. 
It is fixed by the braneworld background and therefore inherits any small deformation of the bulk matter profile. 
Such deformations are expected, for example, from internal-structure dynamics of the brane or from weak couplings to additional bulk fields~\cite{Csaki:2000fc,Zhu:2024gvl,Chen:2020zzs,Campos:2001pr,Bazeia:2004dh,Dzhunushaliev:2006vv,Cruz:2013uwa}. 
To keep the analysis as model-independent as possible, we parameterize the perturbation through a deformed superpotential $W(z)=W_0(z)+\delta W(z)$ and study the two representative profiles shown in Fig.~\ref{figure-1}: a near-brane deformation (Type I) and a distant deformation (Type II). 
Their explicit expressions are collected in Supporting Information. 
Here $\epsilon$ is the deformation amplitude, $w$ controls the width of the Type I near-brane structure, and $a$ sets the location of the Type II remote structures.
Type I produces an additional shallow well close to the original barrier, whereas Type II introduces weak structures centered far from the brane.

The construction should be read as a controlled effective-potential-level parameterization motivated by explicit Einstein--scalar background calculations. 
As shown in Supporting Information, one can perturb the bulk scalar field, reconstruct the warp factor from the Einstein equations, and obtain smooth near-brane or far-brane deformations of the tensor potential while keeping the graviton zero mode localized. 
The analytic Type I and Type II profiles used below are simplified representatives of these numerically generated patterns; we do not claim that each chosen $\delta W$ admits a unique closed-form Einstein--scalar completion.

These ans\"atze are intended to isolate the physical role of perturbation locality rather than to exhaust possible microscopic realizations. 
Near-brane deformations mimic situations in which the internal brane profile is slightly reshaped exactly where the graviton wave functions have the largest support. 
Far-brane deformations probe the opposite regime, where the local geometry around the brane is almost untouched but the asymptotic scattering environment is modified. 
Comparing these two limits lets us separate local control of the prompt ringdown from delayed sensitivity stored in multiple reflections and echoes.

Within this parameterized family, the tensor operator remains factorized and the large-$|z|$ asymptotics of $W$ continue to yield a square-integrable zero mode throughout the parameter range studied, as shown in Supporting Information. 
This localization condition is central to the construction: four-dimensional gravity is preserved while the massive sector is allowed to reorganize.
The question is then sharp and self-contained: once localization is protected, how unstable can the massive QNM spectrum become?

\section{Results}
\subsection{Frequency domain: spectral butterfly effect}
We compute QNM spectra for Type I (near-brane) and Type II (far-brane) deformations 
using the Bernstein spectral method~\cite{Fortuna:2020obg} and the shooting method~\cite{Pani:2013pma}, 
respectively; see Supporting Information for details.
Throughout, the ``fundamental mode'' refers to the least-damped QNM, distinct from the protected graviton zero mode.
The goal is not to construct the full pseudospectrum of the braneworld operator, but to test whether physically mild localized deformations already trigger nonperturbative spectral motion. 
A full pseudospectral treatment requires a separate choice of operator domain, compactification, and norm for the outgoing non-self-adjoint problem. 
Here we therefore adopt a more limited but directly constructive test: we perturb the potential by explicit smooth profiles and track the resulting QNM migration, leaving the full pseudospectral construction as a separate problem.

This distinction is important. 
The graviton zero mode remains protected by localization, whereas the role of the dominant ringdown channel can change after the perturbation. 
Accordingly, we distinguish the deformed first branch $m_1^{(\epsilon)}$ from the least-damped mode $\bar{m}^{(\epsilon)}$, because for sufficiently strong spectral migration they no longer coincide.

\begin{figure*}[!t]
    \centering
    \includegraphics[width=\textwidth]{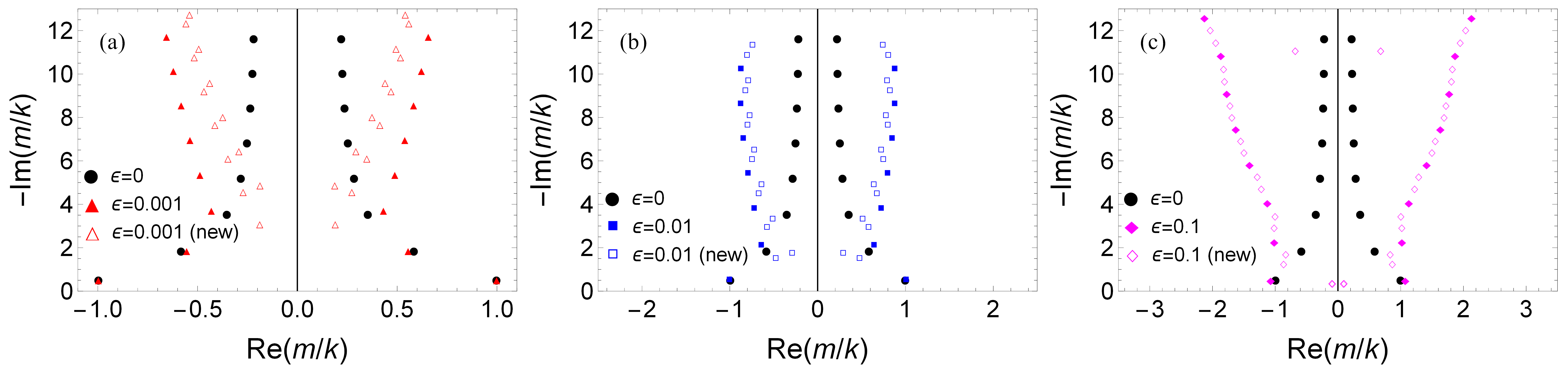}
    \caption{Migration of the QNM spectrum under Type I perturbations with $w=3$: 
    (a) $\epsilon=10^{-3}$, (b) $\epsilon=10^{-2}$, and (c) $\epsilon=10^{-1}$. 
    Filled symbols denote the original QNM branches, while open symbols denote newly generated modes.}
    \label{figure-2}
\end{figure*}

The near-brane deformation reveals the first mechanism of instability. 
As shown in Fig.~\ref{figure-2}, increasing $\epsilon$ does not merely shift the original QNM branches. 
The shallow well generated by Type I perturbations nucleates new pairs of modes that gradually move toward the unperturbed spectrum. 
The least-damped branch changes only moderately, but the higher overtones become increasingly sensitive and may be overtaken by newly generated states. 
The spectral response is therefore not a uniform perturbative drift; it is a qualitative rearrangement of the upper part of the spectrum.

The part of the spectrum most relevant to observation is therefore not the most fragile. 
The first instability appears higher up in the overtone tower, where the modes are more weakly trapped and therefore more vulnerable to a local reshaping of the barrier. 
This hierarchy already hints that frequency-domain fragility need not translate directly into an equally dramatic time-domain signal.

\begin{figure*}[!t]
    \centering
    \includegraphics[width=0.82\textwidth]{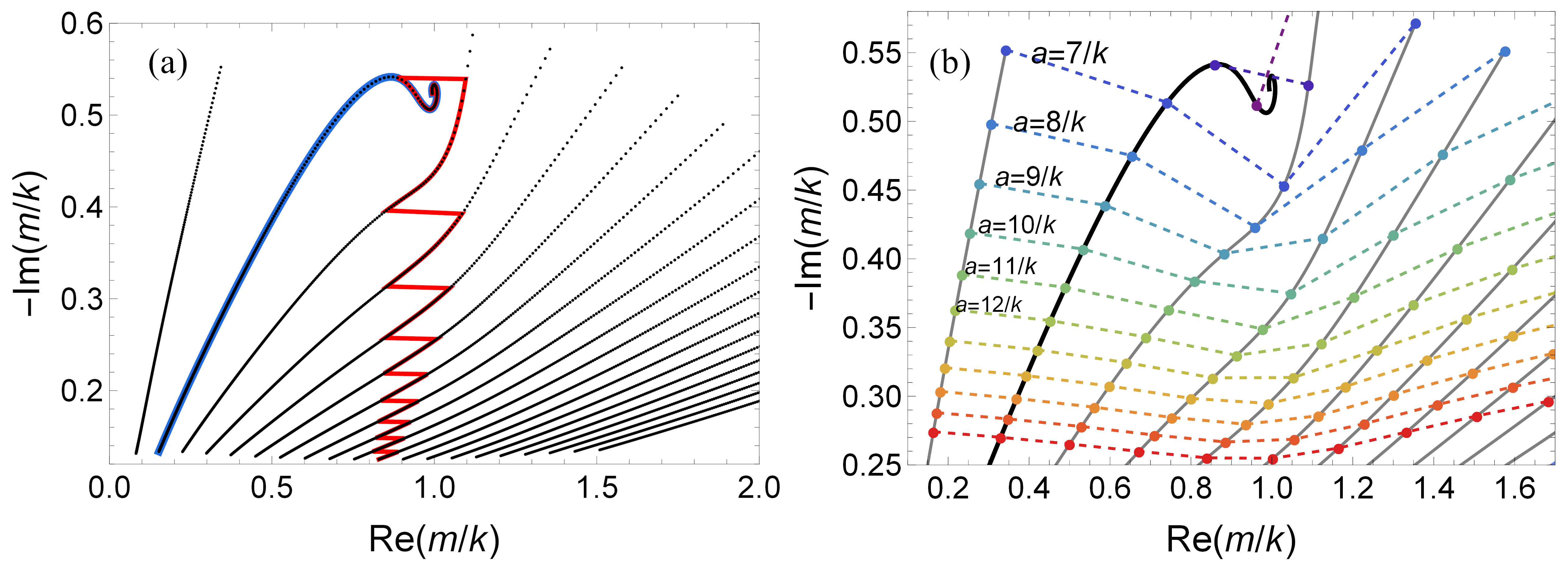}
    \caption{Migration of the QNMs for Type II perturbations with $\epsilon=10^{-3}$. 
    (a) QNM trajectories as a function of $a$; the blue and red curves track, respectively, the original fundamental mode and the least-damped mode after branch switching. 
    (b) Contour representation in the complex-frequency plane; modes with the same $a$ share the same color and are connected by dotted lines.}
    \label{figure-3}
\end{figure*}

\begin{figure*}[!t]
    \centering
    \includegraphics[width=\textwidth]{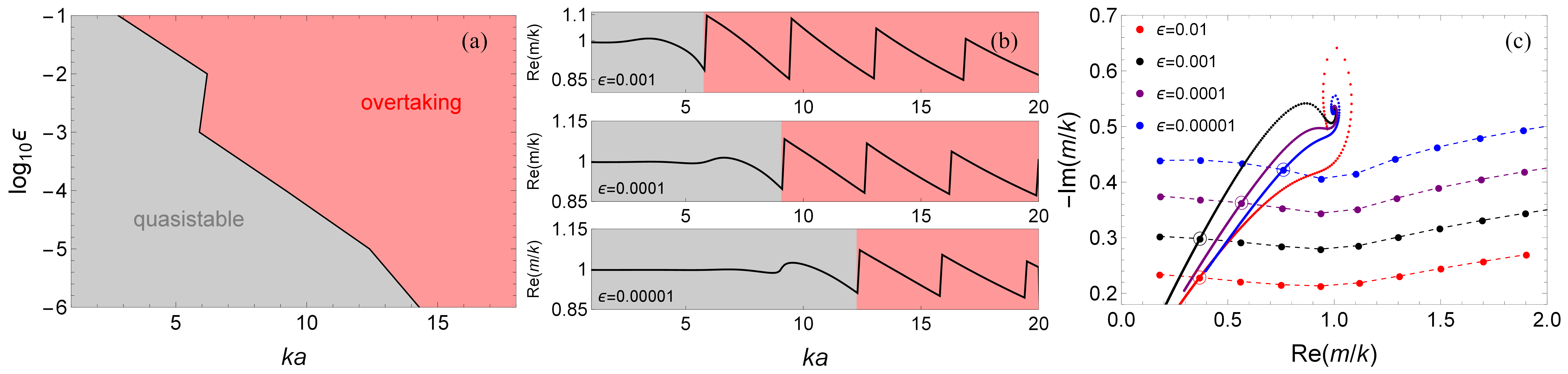}
    \caption{Type II spectral migration and mode overtaking. 
    (a) Phase diagram in the $(ka,\log_{10}\epsilon)$ plane; gray and red denote the quasistable and overtaking regions, respectively. 
    (b) Real part of the least-damped mode $\bar{m}^{(\epsilon)}$ as a function of $ka$ for $\epsilon=10^{-3}$, $10^{-4}$, and $10^{-5}$ from top to bottom.
    (c) Migration of the original fundamental mode and nearby QNMs for $a=15/k$; the colors distinguish different values of $\epsilon$.}
    \label{figure-4}
\end{figure*}

The distant deformation produces a more dramatic effect. 
Figure~\ref{figure-3} shows that, as the perturbation is pushed away from the brane, the original fundamental mode first executes a clockwise spiral in the complex plane and then drifts toward the origin. 
At the same time, new branches enter the spectrum and compete to become the least-damped mode. 
Denoting the perturbed first QNM by $m_1^{(\epsilon)}$, we find in the spiral regime that
\begin{equation}
    \left\lvert \frac{m_1^{(\epsilon)}-m_1^{(0)}}{m_1^{(0)}} \right\rvert \gg \epsilon,
\end{equation}
so the spectral displacement can be parametrically larger than the perturbation itself. 
Figure~\ref{figure-4} makes the transition transparent: the system crosses from a quasistable region, where the original branch remains the least damped, to an overtaking region, where a newly generated mode replaces it as the physically dominant mode. 
The real part of the least-damped mode stays close to the unperturbed value, whereas the imaginary part steadily approaches zero.

The phase diagram also clarifies the scaling of the effect. 
As $\epsilon$ decreases, the critical distance required for overtaking moves outward, reflecting the reduced trapping efficiency of a weaker remote deformation. 
Once the system enters the overtaking regime, however, the dominant long-lived mode is set mainly by the effective cavity formed between the original barrier and the distant structure, consistent with the echo picture discussed below, rather than by the local perturbation amplitude alone.
This is more than the generic statement that a remote weak barrier can eventually generate echoes. 
The nontrivial point is the fate of the least-damped QNM itself: an arbitrarily small distant deformation can first drive the original branch along a spiral trajectory; once the cavity size exceeds a critical threshold, that branch is overtaken by a different, longer-lived one.
This behavior has a natural interpretation in the language of non-Hermitian open systems. 
The outgoing boundary condition makes the QNM problem non-self-adjoint, and the Type II deformation generates a new long-lived branch associated with the effective cavity between the central barrier and the remote structure. 
The observed branch switching and overtaking can therefore be viewed as competition between the original QNM branch and this newly generated branch. 
The spiral motion also suggests that exceptional-point-like structures may exist in the complexified parameter space, although locating them requires a dedicated pseudospectral analysis beyond the scope of the present work.

Taken together, Figs.~\ref{figure-2}--\ref{figure-4} establish the spectral butterfly effect in thick braneworlds. 
Near-brane deformations primarily destabilize higher overtones, while distant deformations can destabilize the least-damped branch itself through spiral migration, branch switching, and overtaking. 
In both cases, a minute perturbation of the potential leads to a macroscopic rearrangement of the spectral fingerprint, even though the detailed mechanism depends strongly on where the deformation is placed.

\subsection{Time domain: resilient ringdown}
To determine which part of this spectral rearrangement is actually visible, we evolve odd Gaussian wave packets in the time domain. 
The numerical integration, extraction strategy, and fitting-window checks are described in Supporting Information. 
The odd initial data deliberately suppress the even zero mode and isolate the QNM response, which makes the contrast between prompt ringdown and late-time echoes especially transparent. 
This diagnostic choice allows us to identify which parts of the spectral instability enter the waveform before adding source-dependent astrophysical complications.
For more general initial data, an even component would also excite the localized zero mode and continuum or tail contributions. 
These components would change the slowly varying part of the waveform and can make late echoes harder to isolate, 
but they do not eliminate the underlying QNM branches themselves. 
Thus the statements about prompt-ringdown resilience should be read as statements about the QNM-dominated diagnostic channel, not as claims about every possible source profile.

For numerical convenience, we work in light-cone coordinates, where the wave equation takes the form
\begin{equation}
    4 \partial_u \partial_v \Psi(u,v) + V(u,v)\Psi(u,v)=0.
\end{equation}
This representation is useful because it separates the prompt signal generated near the brane from the delayed contributions created when the wave packet repeatedly scatters off distant structures.

\begin{figure*}[htb]
    \centering
    \includegraphics[width=0.94\textwidth]{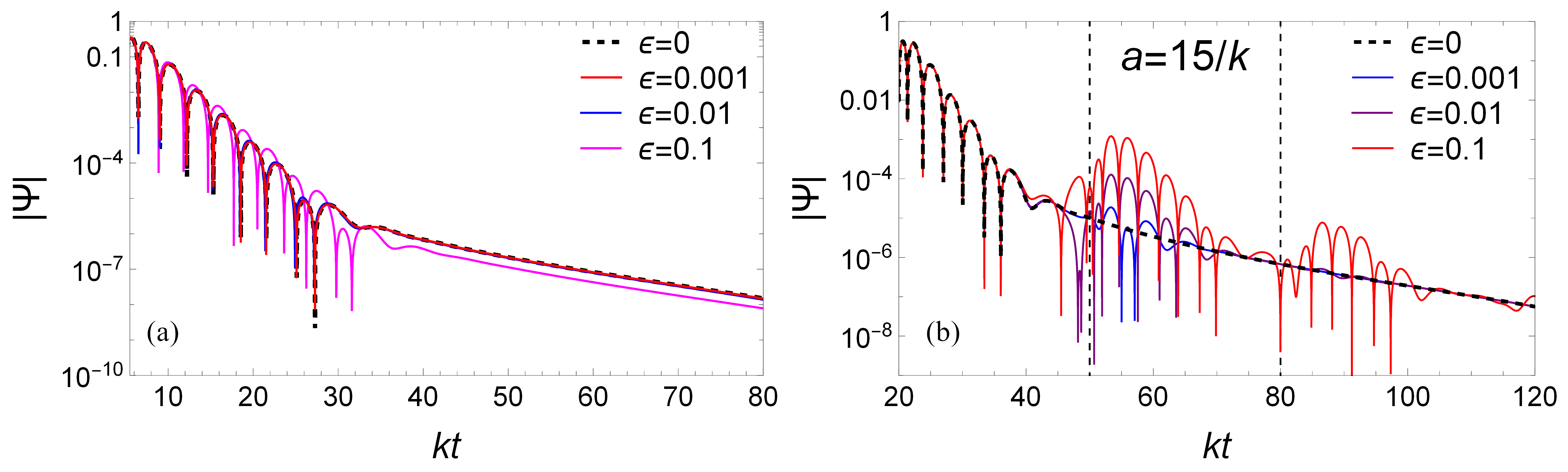}
    \caption{Time-domain waveforms. 
    (a) Type I perturbations with several values of $\epsilon$ and $w=3$, extracted at $kz_{\text{ext}}=5$. 
    (b) Type II perturbations with several values of $\epsilon$ and $a=15/k$, extracted at $kz_{\text{ext}}=20$. 
    The vertical dashed lines in panel (b) mark the expected first and second echo arrivals at $t=t_{\mathrm{peak}}+2a$ and $t=t_{\mathrm{peak}}+4a$, respectively.}
    \label{figure-5}
\end{figure*}

Figure~\ref{figure-5}(a) shows the response to Type I perturbations. 
The prompt signal is clearly modified: the oscillation frequency shifts with $\epsilon$, and the early ringdown follows the local deformation of the potential near the brane. 
Equally important is the absence of a comparably dramatic imprint in the waveform. 
Although the frequency-domain spectrum shows strong overtone sensitivity, the finite-time waveform is still dominated by the original fundamental branch. 
The overtone instability is therefore largely hidden in the early-time ringdown.

Figure~\ref{figure-5}(b) shows the complementary behavior for Type II perturbations. 
As the deformation is moved away from the brane, its impact on the prompt ringdown becomes weaker because the early waveform is controlled mainly by the geometry in the brane neighborhood. 
At late times, however, clear echoes emerge. 
We define $\Delta t=t-t_{\mathrm{peak}}$, where $t_{\mathrm{peak}}$ is the peak time of the initially arriving burst.
They appear at delays $\Delta t \simeq 2a, 4a, \ldots$, exactly as expected from repeated partial reflections between the original barrier and the distant perturbation. 
Some of these echoes decay more slowly than the prompt signal, in agreement with the appearance of a new least-damped branch in the frequency-domain analysis.

This delayed structure is the time-domain counterpart of the branch switching seen in the frequency domain. 
The prompt waveform is still governed by the original barrier because the initial packet encounters that barrier first. 
Only after one or more round trips through the enlarged scattering region can the distant deformation reorganize the signal, which shifts the clearest imprint of the spectral butterfly effect into the low-amplitude echo sector rather than the earliest observable cycles. 
As summarized in Supporting Information, this separation between prompt ringdown and late echoes remains robust against variations in numerical resolution, extraction point, and fitting window.

The time-domain analysis therefore clarifies the apparent tension. 
The QNM spectrum is mathematically fragile, but the waveform is not uniformly so. 
Near-brane deformations modify the cleanest part of the signal, yet even there the prompt ringdown remains governed by the original fundamental mode. 
Far-brane deformations generate the dramatic spectral migration seen in Figs.~\ref{figure-3} and~\ref{figure-4}, but their cleanest imprint is postponed into low-amplitude late-time echoes. 
This separation between prompt ringdown and late echoes is the precise sense in which thick braneworlds exhibit a resilient ringdown despite a fragile spectrum.
The waveforms shown here are dimensionless responses of an idealized one-dimensional scattering problem. 
A direct comparison with LIGO/Virgo/KAGRA noise curves would require a source model, detector response, and parameter-estimation pipeline. Such a comparison is beyond the scope of this article. 
Within our normalization, the first Type II echo is suppressed by one additional reflection/transmission through the distant structure and is of order $\epsilon$ relative to the prompt response, while subsequent echoes are smaller. 
Thus small perturbations with $\epsilon=10^{-3}$--$10^{-2}$ naturally place the cleanest spectral-instability signature 
in a low-amplitude part of the waveform, which is not in conflict with current null searches for echoes~\cite{Uchikata:2023}.

\subsection{Frequency-domain/time-domain cross-check}

\begin{figure*}[t]
    \centering
    \includegraphics[width=0.94\textwidth]{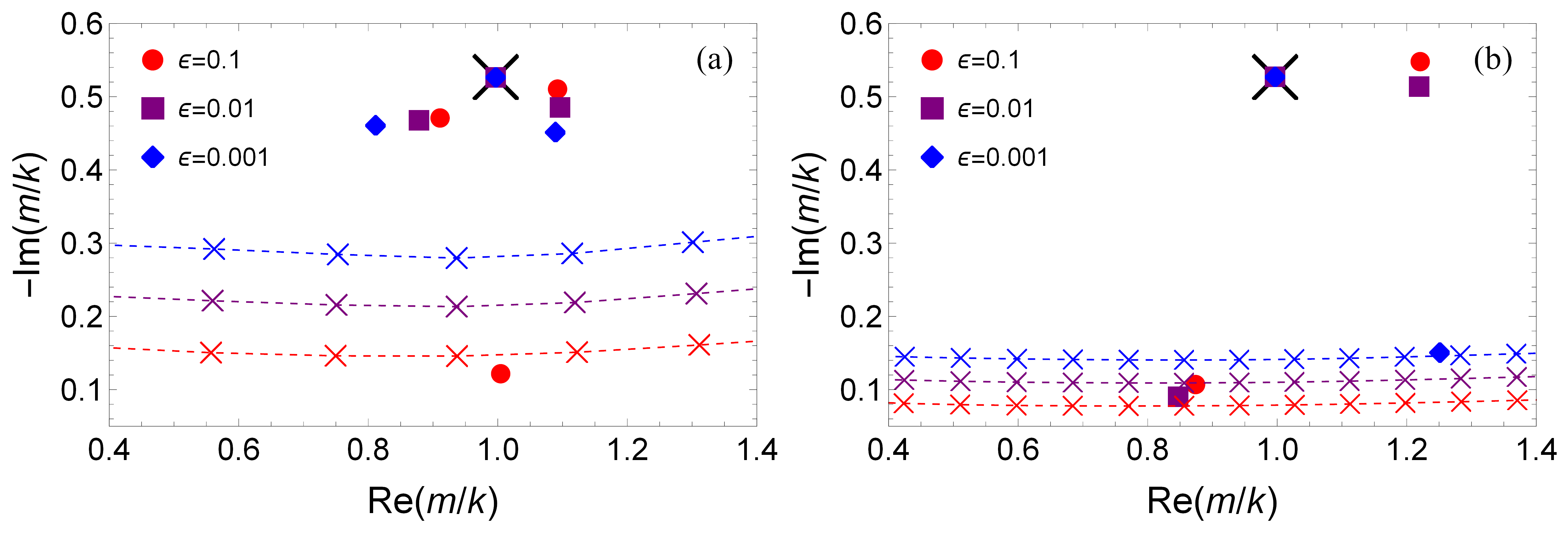}
\caption{Frequency-domain/time-domain comparison for Type II perturbations. 
Crosses denote QNMs obtained by the shooting method, while dots denote damped-sinusoid fits to time-domain waveform windows. 
In the echo windows, several dots can arise from one window because $K=3$ or $K=4$ damped components are retained in the multi-component fit.
(a) $a=15/k$ and (b) $a=35/k$.}
\label{figure-fd-td-typeII-main}
\end{figure*}

\begin{figure*}[t]
    \centering
    \includegraphics[width=0.94\textwidth]{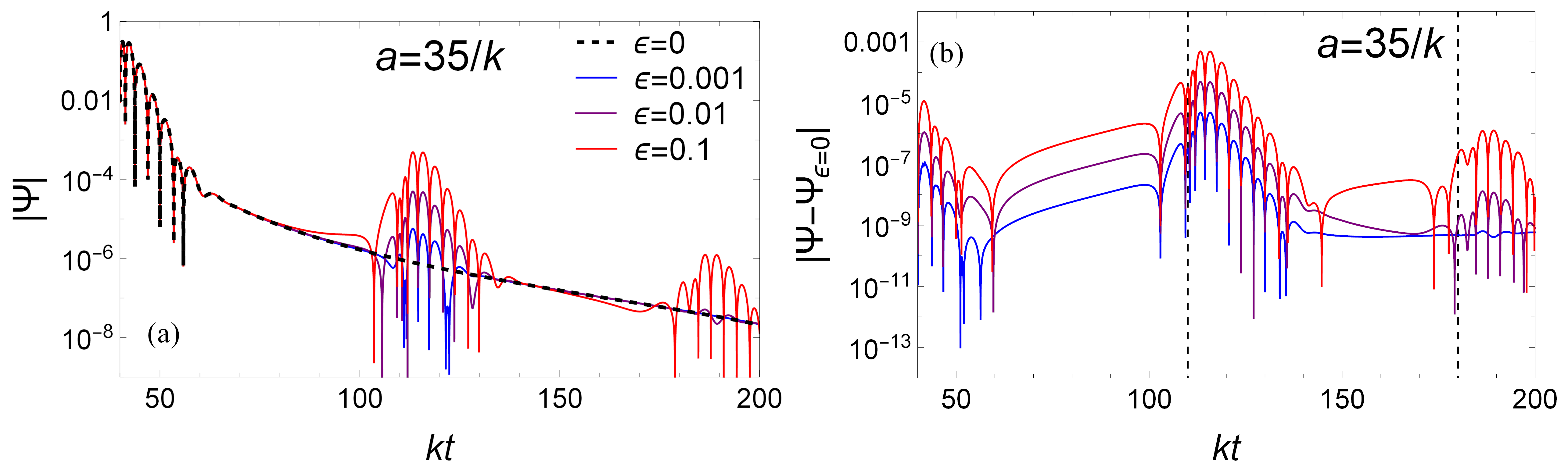}
\caption{Late-time Type II waveform for a remote deformation at $a=35/k$, extracted at $kz_{\rm ext}=40$. 
(a) Absolute value of the waveform for several perturbation strengths. 
(b) Absolute value of the difference from the unperturbed waveform. 
The vertical dashed lines mark the expected first and second echo arrivals at $t=t_{\mathrm{peak}}+2a$ and $t=t_{\mathrm{peak}}+4a$, corresponding to delays $\Delta t\simeq 2a$ and $4a$ relative to the prompt peak.}
\label{figure-typeII-a35-main}
\end{figure*}

We further compare the frequency-domain QNMs with the frequencies extracted directly from the time-domain signals. 
This comparison is important because the spectral butterfly effect is diagnosed in the complex-frequency plane, whereas an observable waveform contains only a finite time interval contaminated by prompt propagation, tails, and numerical noise. 
To extract the QNM frequencies, we fit the waveform by
\begin{equation}
    \Psi(t)=\sum_{\ell=1}^{K} A_{\ell} e^{\operatorname{Im}(\omega_{\ell})t}
    \sin\left[\operatorname{Re}(\omega_{\ell})t-\theta_{\ell}\right],
    \label{fit-timedomain-main}
\end{equation}
where $K$ is the number of fitted damped sinusoids. 
For the prompt ringdown we use $K=1$ to extract the dominant visible mode and repeat selected fits with $K=2$ as a stability check. 
For the Type II echo windows, where several damped components can overlap, we use multi-component fits with $K=3$ or $K=4$; the multiple time-domain points in Fig.~\ref{figure-fd-td-typeII-main} are the individual fitted components retained from these echo-window fits.
The four-dimensional momentum is set to $p=0$ in the present comparison, so the frequency-domain mass parameter and the fitted frequency are identified through $m_n^2=\omega_n^2-p^2$. 
The fitting windows are chosen after the initial burst has peaked and before tail contamination dominates: Type I fits use $k\Delta t\in(10,20)$, Type II prompt fits are restricted to the pre-echo interval $k\Delta t<2ka$, and delayed echoes are fitted in separate windows centered near $\Delta t\simeq2a,4a,\ldots$.
The mismatch is quantified by
\begin{equation}
    \Delta_{\mathrm{rel}}=\frac{\left|\omega_{\mathrm{TD}}-m_{\mathrm{FD}}\right|}{\left|m_{\mathrm{FD}}\right|}.
    \label{rel-diff-main}
\end{equation}
Here the subscripts $\mathrm{TD}$ and $\mathrm{FD}$ denote time-domain and frequency-domain quantities, respectively.
The Type I comparison is summarized in Table~\ref{table-fd-td-typeI-main}. 
The Type II frequency-domain/time-domain comparison and the corresponding late-time waveform are shown in Figs.~\ref{figure-fd-td-typeII-main} and~\ref{figure-typeII-a35-main}, respectively.

\begin{table*}[t]
\centering
\footnotesize
\renewcommand\arraystretch{1.15}
\begin{tabular}{c c c c}
\hline
$\epsilon$ & $\omega_{\rm TD}/k$ & $m_{1,\rm FD}/k$ & $\Delta_{\rm rel}$\\
\hline
$0$       & $0.997038-0.525830i$ & $0.997018-0.526365i$ & $4.75\times10^{-4}$\\
$10^{-3}$ & $0.997644-0.525118i$ & $0.997887-0.525925i$ & $7.47\times10^{-4}$\\
$10^{-2}$ & $1.005590-0.521809i$ & $1.005560-0.522054i$ & $2.18\times10^{-4}$\\
$10^{-1}$ & $1.072130-0.489705i$ & $1.072260-0.489727i$ & $1.12\times10^{-4}$\\
\hline
\end{tabular}
\caption{Frequency-domain/time-domain comparison for Type I perturbations with $w=3$. 
The time-domain values are obtained by fitting Eq.~\eqref{fit-timedomain-main} in the window $k\Delta t\in(10,20)$, while each frequency-domain value is the first QNM computed by the Bernstein spectral method for the corresponding value of $\epsilon$.}
\label{table-fd-td-typeI-main}
\end{table*}

For Type I perturbations, Table~\ref{table-fd-td-typeI-main} shows that the fitted prompt-ringdown frequency agrees with the first frequency-domain QNM at the level of $10^{-4}$--$10^{-3}$. 
Thus the early-time waveform does track the small shift of the original fundamental branch. 
At the same time, the much larger migration of higher overtones is not visible with comparable clarity in the finite-time waveform, because these modes are strongly suppressed by their larger damping rates and by the limited fitting window.

For Type II perturbations, Fig.~\ref{figure-fd-td-typeII-main} provides the complementary check. 
The fitted prompt-ringdown points remain close to the unperturbed fundamental mode, confirming that a distant localized deformation has little direct influence on the first few cycles. 
The multiple late-time fitted points instead arise from echoes and newly generated long-lived components; when the perturbation is moved farther from the brane, these late-time components become more prominent and can have smaller damping rates than the original branch. 
This delayed channel is displayed directly in Fig.~\ref{figure-typeII-a35-main}. 
For $a=35/k$, the prompt waveform almost overlaps with the unperturbed signal, while the difference plot isolates echo peaks near the expected round-trip times $\Delta t\simeq 2a$ and $4a$. 
The echo sector therefore carries the frequency-domain branch competition more efficiently than the earliest ringdown cycles. 
This direct comparison is the numerical basis for our conclusion that frequency-domain fragility and prompt-ringdown resilience coexist: the spectrum is reorganized, but the large overtone migration remains largely hidden in the prompt waveform, whereas branch competition is revealed most clearly through low-amplitude delayed echoes rather than through the earliest ringdown cycles.

\section{Discussion and conclusion}
For a canonical thick brane under smooth localization-preserving deformations, we find a clear separation between spectral sensitivity and waveform sensitivity.
Near-brane reshaping destabilizes higher overtones, whereas remote deformations can drive spiral migration, branch switching, and overtaking of the least-damped mode. 
Throughout, the graviton zero mode remains localized, so four-dimensional gravity is preserved while the massive spectrum reorganizes.

The time-domain response is more selective than the full spectrum. 
In the diagnostic setup used here, the prompt ringdown continues to be controlled mainly by the original barrier, while the clearest imprint of distant deformations is displaced into delayed echoes. 
The strongest spectral instabilities therefore appear first where finite-time waveforms have the lowest immediate sensitivity: in higher overtones, mode competition, and low-amplitude late-time structure.

The broader lesson is that frequency-domain fragility and finite-time waveform observability are distinct aspects of braneworld spectroscopy: the QNM spectrum can undergo order-one rearrangements while the prompt ringdown remains controlled by the original barrier. 
This coexistence is demonstrated explicitly for one canonical thick brane and two representative localized deformation families. 
Its relevance is broader within the class of Einstein--matter thick branes: matter or geometric perturbations are encoded in the warp factor and hence in the tensor superpotential $W$, and the canonical calculation shows that the induced tensor-potential responses can be organized into near-brane reshaping and far-brane remote-structure formation. 
Thus the method and qualitative conclusions apply to localization-preserving Einstein--matter thick branes at the level of these two effective-potential response classes. 
For thick brane models in modified gravity or other non-Einstein--matter frameworks, the qualitative conclusions established here should not be assumed to hold without a separate, model-specific calculation; nevertheless, the present deformation-and-tracking method provides a direct way to study them.
This perspective sharpens the KK fingerprint by showing where it is stable, where it is hidden, and where late-time echoes and mode competition can reveal otherwise invisible bulk structure.

\section*{Conflict of interest}

The authors declare that they have no conflict of interest.

\section*{Acknowledgments}

We would like to thank Qin Tan and Wen-Yi Zhou for very useful discussions. 
This work was supported by 
the National Natural Science Foundation of China (Grant Nos. 12475056, 12575055, 12205129, and 12247101), 
Gansu Province's Top Leading Talent Support Plan,
the Fundamental Research Funds for the Central Universities (Grant Nos. lzujbky-2025-it05 and lzujbky-2025-jdzx07), 
the Natural Science Foundation of Gansu Province (Grant Nos. 22JR5RA389 and 25JRRA799), 
and the `111 Center' under Grant No. B20063. 
Wen-Di Guo and Yu-Peng Zhang were supported by the ``Talent Scientific Fund of Lanzhou University''.

\setcounter{section}{0}
\setcounter{equation}{0}
\setcounter{figure}{0}
\renewcommand{\thesection}{\Alph{section}}
\renewcommand{\thefigure}{S\arabic{figure}}
\renewcommand{\theequation}{S\arabic{equation}}
\renewcommand{\theHsection}{supp.\arabic{section}}
\renewcommand{\theHfigure}{supp.\arabic{figure}}
\renewcommand{\theHequation}{supp.\arabic{equation}}

\section*{\centering Supporting Information}

\section{Representative perturbations and their scope}

The article studies \emph{representative parameterized deformations} of the tensor potential 
designed to isolate two broad spectral-response mechanisms in Einstein--matter thick branes: 
local barrier reshaping near the brane and remote cavity formation far from the brane.
The physical requirement imposed throughout is that the deformed potential must still support a normalizable graviton zero mode, 
so that four-dimensional gravity is preserved while the massive spectrum is allowed to reorganize.
This section clarifies what is, and is not, assumed about the five-dimensional background. 
The Type I and Type II deformations used in the main text are not presented as a complete classification of all exact back-reacted thick brane solutions. 
They are analytic representatives of potential deformations motivated by explicit Einstein--scalar numerical backgrounds. 
For a general Einstein--matter thick brane, perturbations of the matter sector or of the background geometry are encoded in a change of the warp factor and therefore in the tensor superpotential $W=-3\partial_z A/2$ and the tensor effective potential. 
The explicit canonical calculation below shows that such effective-potential responses can be organized into two broad classes, near-brane reshaping and far-brane remote-structure formation, which are then represented analytically by the Type I and Type II profiles. 
In this sense, the method and qualitative conclusions of the article are expected to apply to localization-preserving Einstein--matter thick branes at the level of these two response classes. 
For thick brane models in modified gravity or other non-Einstein--matter frameworks, these qualitative conclusions should not be assumed to hold without a separate, model-specific calculation; nevertheless, the deformation-and-tracking strategy can be extended to analyze them.

For the canonical thick brane used in the main text, the representative deformations are motivated by the standard five-dimensional Einstein--scalar system~\cite{DeWolfe:1999cp,Gremm:1999pj}:
\begin{equation}
    S=\int d^4x\,dy\,\sqrt{-g}\left[\frac{1}{2}R-\frac{1}{2}\nabla_M\phi\nabla^M\phi-U(\phi)\right],
\end{equation}
where we have set the five-dimensional gravitational constant to unity. 
The background is assumed to preserve four-dimensional Poincar\'e invariance, with the metric ansatz
\begin{equation}
    ds_5^2=e^{2A(y)}\eta_{\mu\nu}dx^\mu dx^\nu+dy^2,
\end{equation}
and with both the warp factor $A$ and the bulk scalar $\phi$ depending only on the extra-dimensional coordinate $y$. 
The background equations are
\begin{align}
    \phi''+4A'\phi'&=\frac{\partial U}{\partial \phi},\\
    6A'^2+3A''&=-\frac{1}{2}\phi'^2-U(\phi),\\
    6A'^2&=\frac{1}{2}\phi'^2-U(\phi).
\end{align}
Combining the last two equations gives
\begin{equation}
    3A''(y)=-\phi'^2(y)\leq 0,
\end{equation}
which is the reconstruction equation used below once a perturbed scalar profile is specified.
For the canonical one-parameter thick brane family used to motivate the numerical backgrounds, one may write
\begin{equation}
    A_0(y)=b\ln[\operatorname{sech}(ky)].
\end{equation}
The corresponding scalar profile and scalar potential are
\begin{align}
    \phi_0(y)&=\sqrt{3b}\arcsin[\tanh(ky)],\\
    U(\phi)&=\frac{3bk^2}{4}\left[
    1-4b+(1+4b)\cos\left(\frac{2\phi}{\sqrt{3b}}\right)
    \right].
\end{align}
Here $k$ sets the characteristic energy scale of the five-dimensional bulk, while $b>0$ controls the shape of the warp factor.
Thus the starting point is an exact solution of the five-dimensional Einstein--scalar equations, not a freely prescribed scattering potential.
In the quasinormal-mode (QNM) analysis of the article we set $b=1$, which gives the explicit background used in the main text.
Passing to the conformal coordinate $z$ through $dz=e^{-A(y)}dy$, this background becomes
\begin{equation}
    A_0(z)=-\frac{1}{2}\ln(1+k^2 z^2).
\end{equation}
For transverse-traceless tensor perturbations, the metric is written as
\begin{equation}
    ds^2=e^{2A(z)}\left[(\eta_{\mu\nu}+h_{\mu\nu})dx^\mu dx^\nu+dz^2\right],
\end{equation}
and the corresponding mode equation can be recast into the Schr\"odinger form
\begin{equation}
    \left[-\partial_z^2+V_0(z)\right]\psi(z)=m^2\psi(z),
\end{equation}
with
\begin{equation}
    V_0(z)=\frac{3}{2}\partial_z^2A_0+\frac{9}{4}(\partial_zA_0)^2,
\end{equation}
where $m^2=\omega^2-p^2$. 
Here $\omega$ and $p$ denote the four-dimensional frequency and spatial momentum, respectively.
At this stage it is natural to introduce the superpotential
\begin{align}
    W_0(z)&=-\frac{3}{2}\partial_z A_0(z)=\frac{3k^2 z}{2(1+k^2 z^2)},\\
    V_0(z)&=W_0^2(z)-\partial_z W_0(z)
    =\frac{3k^2\left(5k^2z^2 - 2\right)}{4 \left(k^2z^2 +1\right)^2},
\end{align}
so that the tensor operator is manifestly factorized and the undeformed zero mode is
\begin{equation}
    \psi_0^{(0)}(z)\propto e^{\frac{3}{2}A_0(z)}=(1+k^2 z^2)^{-3/4}.
\end{equation}

The deformations induced by matter-field fluctuations can then be described by writing
\begin{equation}
    W(z)=W_0(z)+\delta W(z),\qquad V(z)=W^2(z)-\partial_z W(z),
\end{equation}
or equivalently
\begin{align}
    \delta V(z)&=V(z)-V_0(z)\nonumber\\
    &=2W_0(z)\delta W(z)+\delta W^2(z)-\partial_z\delta W(z).
\end{align}
This construction makes explicit how a localized perturbation of the superpotential 
generates a localized reshaping of the scattering barrier while keeping the supersymmetric factorization manifest.

\begin{figure*}[htb]
    \centering
    \includegraphics[width=0.82\textwidth]{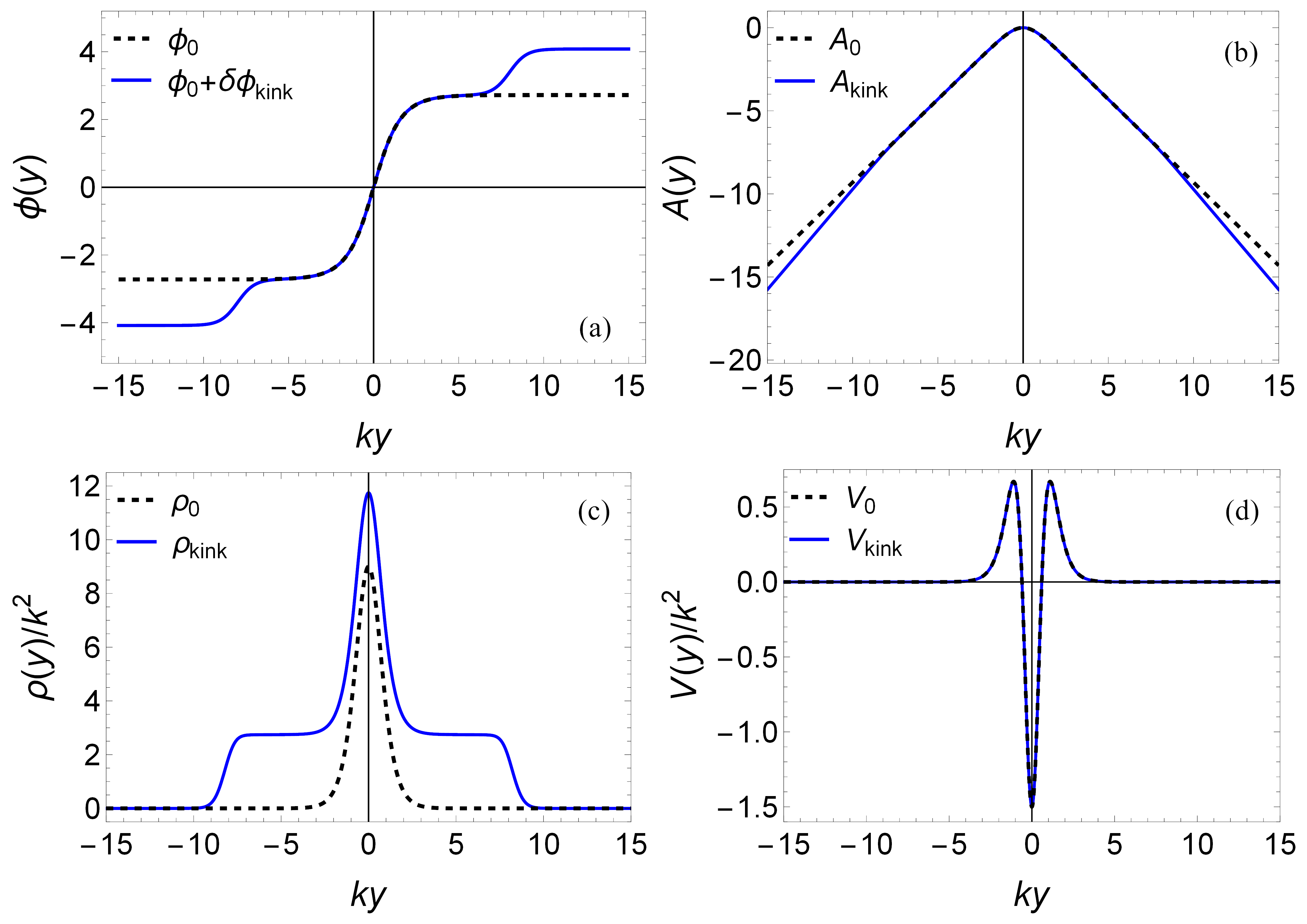}
    \caption{Einstein--scalar numerical background generated by a double-kink perturbation of the bulk scalar field: 
    (a) the bulk scalar profile, (b) the reconstructed warp factor, (c) the energy density, and (d) the tensor effective potential. 
    Dashed black and solid blue curves denote the unperturbed and perturbed backgrounds, respectively. 
    The warp factor is obtained from $3A''=-(\phi_0'+\delta\phi_{\mathrm{kink}}')^2$, and the energy density and tensor effective potential are then computed from the Einstein equations. 
    The parameters are $b=1$, $\epsilon_m=1/4$, and $a_m=8/k$.}
    \label{figure-bg-deformation}
\end{figure*}

The representative deformations are motivated by the explicit Einstein--scalar background calculations carried out in our detailed analysis. 
There, the bulk scalar field is perturbed first, and the warp factor and tensor potential are then reconstructed from the field equations rather than chosen independently. 
For example, an odd double-kink matter perturbation,
\begin{align}
    \delta\phi_{\mathrm{kink}}(y)=\epsilon_m\frac{\pi\sqrt{3b}}{2}
    &\left\{\tanh[k(y-a_m)]\right.\nonumber\\
    &\left.+\tanh[k(y+a_m)]\right\},
\end{align}
keeps the induced warp factor and tensor effective potential even. 
Here $\epsilon_m$ and $a_m$ denote the amplitude of the matter perturbation and its displacement from the brane, respectively.
Substituting $\phi=\phi_0+\delta\phi_{\mathrm{kink}}$ into the Einstein--scalar constraint gives
\begin{equation}
    3A''(y)=-\left[\phi_0'(y)+\delta\phi_{\mathrm{kink}}'(y)\right]^2,
\end{equation}
which can be solved numerically with $A(0)=A'(0)=0$. 
The energy density and the tensor potential are then computed from
\begin{align}
    \rho(y)&=-\left(6A'^2+3A''\right),\\
    V(y)&=e^{2A}\left(\frac{3}{2}A''+\frac{15}{4}A'^2\right).
\end{align}
The scalar self-interaction potential supporting the numerical profile can also be reconstructed along the solution curve from
\begin{equation}
    U_{\mathrm{rec}}(y)=\frac{1}{2}\left[\phi_0'(y)+\delta\phi_{\mathrm{kink}}'(y)\right]^2-6A'^2(y).
\end{equation}
When the scalar profile is monotonic, this gives $U_{\mathrm{rec}}(\phi)$ directly; otherwise it gives the required branchwise reconstruction along the background solution. 
This is the concrete sense in which the potential deformation is tied to an Einstein--scalar background.
Figure~\ref{figure-bg-deformation} shows the scalar profile and the reconstructed geometry, energy density, and tensor potential for a representative double-kink perturbation.

\begin{figure*}[htb]
    \centering
    \includegraphics[width=\textwidth]{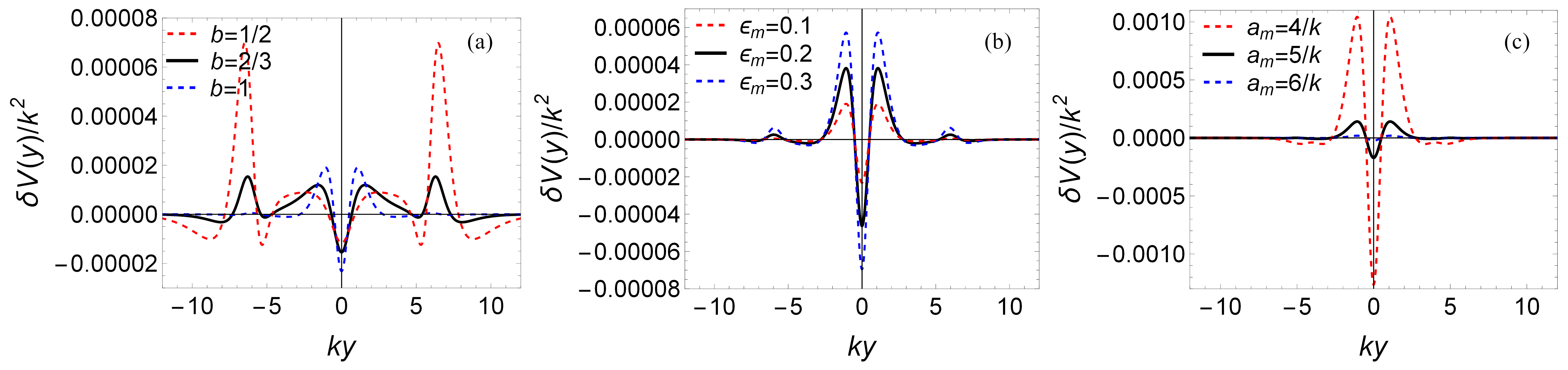}
    \caption{Matter-induced deformation of the tensor effective potential in the numerical Einstein--scalar calculation. 
    (a) Dependence on $b$ for $\epsilon_m=0.1$ and $a_m=6/k$. 
    (b) Dependence on $\epsilon_m$ for $b=1$ and $a_m=6/k$. 
    (c) Dependence on $a_m$ for $b=1$ and $\epsilon_m=0.1$. 
    These results motivate the analytic Type I and Type II representatives used in the article: near-brane reshaping and remote symmetric structures, respectively.}
    \label{figure-bg-deltaV}
\end{figure*}

The numerical Einstein--scalar calculation also shows how the background shape parameter $b$, perturbation position $a_m$, and perturbation strength $\epsilon_m$ control the induced effective potential. 
For the symmetric double-kink family, the energy-density change forms a plateau between the two kinks, while the effective-potential deformation can be concentrated either near the brane or near the remote matter structures depending on $b$, $a_m$, and $\epsilon_m$. 
For $b=1$ and $a_m=6/k$, the detailed calculation gives the indicative hierarchy between the characteristic peak amplitudes
\begin{equation}
    \frac{\max_{y}|\delta\rho(y)|}{k^2}\sim 10\,\epsilon_m,\qquad
    \frac{\max_{y}|\delta V(y)|}{k^2}\sim 10^{-4}\epsilon_m,
\end{equation}
for the symmetric far-brane case, with the precise ratio depending on $a_m$. 
This explains why a very small effective-potential deformation can be physically generated by a much larger but smooth rearrangement of the bulk matter distribution. 
The Type I and Type II profiles used in the article are analytic representatives of these two numerical patterns: 
Type I captures the near-brane potential response, 
while Type II captures the symmetric far-brane response localized near $|y|\simeq a_m$. 
They are therefore not arbitrary potentials, but simplified analytic proxies for Einstein--scalar-induced deformations that preserve the two essential features needed here: smoothness and zero-mode localization.
Figure~\ref{figure-bg-deltaV} illustrates how the induced tensor-potential deformation changes with $b$, $\epsilon_m$, and $a_m$.

The two representative deformations used in the main text are
\begin{align}
    \label{Type-I-superpotential}
    \text{Type I:}\quad \delta W(z)=&\epsilon \frac{w k^2 z\left(k^2 z^2-2w^2\right)^2}{2\left(k^2 z^2+w^2\right)^3},\\
    \label{Type-II-superpotential}
    \text{Type II:}\quad \delta W(z)=&\epsilon \frac{z}{a^2}\left[\frac{2 k^2(z-a)^2-1}{\left[k^2(z-a)^2+1\right]^2}\right. \nonumber \\
    &\left. + \frac{2 k^2(z+a)^2-1}{\left[k^2(z+a)^2+1\right]^2}\right].
\end{align}
In both families, $\epsilon$ is the deformation amplitude.
Both choices of $\delta W$ are odd functions of $z$, so the corresponding deformed potentials remain even. 
Type I provides a near-brane deformation controlled by the width parameter $w$; it primarily reshapes the central barrier and the shallow well around the brane. 
Type II produces a symmetric pair of distant structures centered near $\pm a$; for $ka\gg 1$, it leaves the brane neighborhood almost unchanged while adding a remote scattering region. 
This is the geometric origin of the cavity-like behavior and the echo delay scale $\Delta t \simeq 2a,4a,\ldots$ discussed in the main text.

The two families are chosen for their representativeness: 
they capture the two qualitative situations central to the article, namely local barrier reshaping and remote cavity formation. 
They therefore provide a controlled testbed for the spectral butterfly effect 
while preserving the localized graviton zero mode.

\section{Zero-mode localization}

Because the tensor potential always admits the factorized form
\begin{equation}
    -\partial_z^2+V(z)=\left(-\partial_z+W\right)\left(\partial_z+W\right),
\end{equation}
the graviton zero mode satisfies
\begin{equation}
    \left(\partial_z+W\right)\psi_0(z)=0,
\end{equation}
and therefore
\begin{equation}
    \psi_0(z)\propto \exp\left[-\int^z W(z')\,dz'\right].
\end{equation}
For the parameterized family used here, the most direct way to check localization is to examine the large-$|z|$ asymptotics of $W(z)$.

For Type I one finds
\begin{equation}
    W(z)=\frac{3+\epsilon w}{2z}+\mathcal{O}(z^{-3}),\qquad |z|\to\infty,
\end{equation}
which implies
\begin{equation}
    \psi_{0,\mathrm{I}}(z)\sim |z|^{-(3+\epsilon w)/2}.
\end{equation}
For Type II,
\begin{equation}
    W(z)=\left(\frac{3}{2}+\frac{4\epsilon}{k^2a^2}\right)\frac{1}{z}+\mathcal{O}(z^{-3}),
    \qquad |z|\to\infty,
\end{equation}
so that
\begin{equation}
    \psi_{0,\mathrm{II}}(z)\sim |z|^{-3/2-4\epsilon/(k^2a^2)}.
\end{equation}
Hence, for the parameter ranges considered in the main text, the zero mode decays faster than $|z|^{-1/2}$ in both cases and is therefore square integrable:
\begin{equation}
    \int_{-\infty}^{\infty}\left|\psi_0(z)\right|^2 dz<\infty.
\end{equation}

This asymptotic analysis directly establishes the condition relevant to the present article: 
the representative deformations used to probe the QNM spectrum preserve the localized graviton zero mode and thus retain four-dimensional gravity at low energies.

\begin{figure*}[htb]
    \centering
    \includegraphics[width=0.82\textwidth]{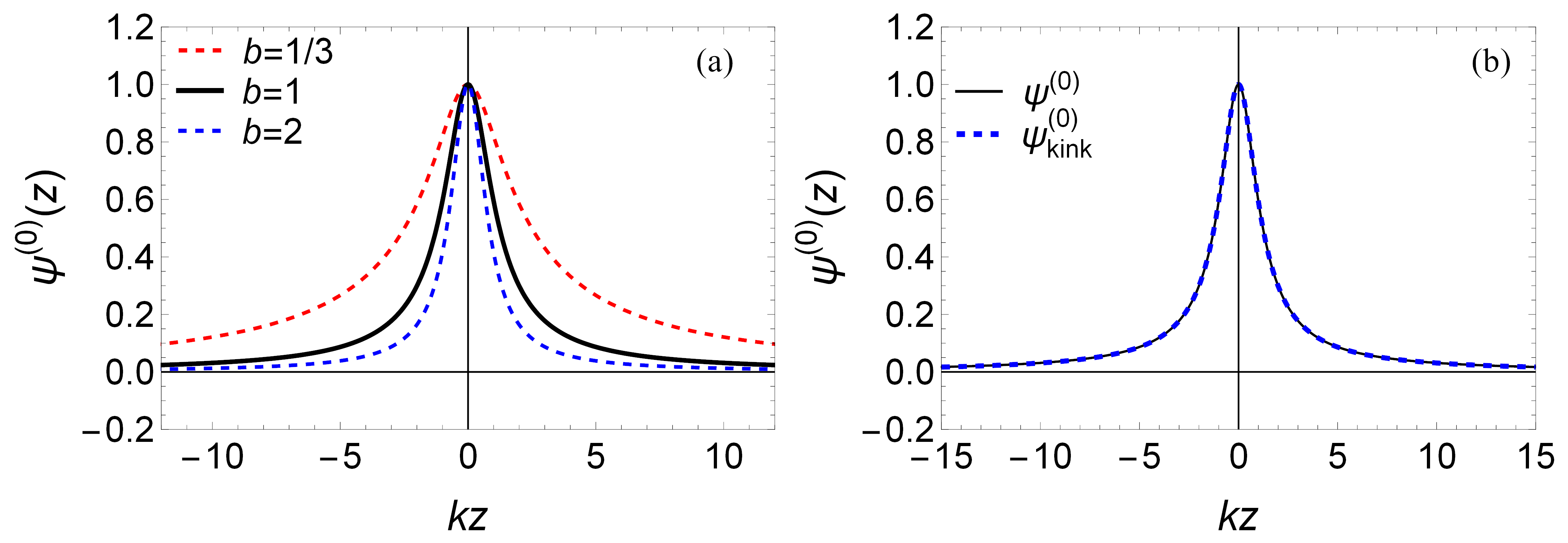}
    \caption{Zero-mode profiles in the Einstein--scalar numerical backgrounds: 
    (a) unperturbed profiles with $\epsilon_m=0$ for $b=1/3$, $1$, and $2$; 
    (b) comparison of the unperturbed profile with a symmetric perturbation for $b=1$, $\epsilon_m=0.1$, and $a_m=4/k$. 
    The perturbation deforms the profile but preserves normalizability.}
    \label{figure-bg-zeromode}
\end{figure*}

The same conclusion is obtained directly in the explicit Einstein--scalar numerical backgrounds discussed in Sec.~A. 
In the conformal coordinate, the Einstein equations imply
\begin{equation}
    \partial_z^2A-\left(\partial_zA\right)^2
    =-\frac{1}{3}\left(\partial_z\phi\right)^2\leq 0.
\end{equation}
Equivalently, for $B(z)=e^{-A(z)}$,
\begin{equation}
    \partial_z^2B(z)\geq 0.
\end{equation}
Together with the asymptotically anti-de Sitter (AdS) behavior of the reconstructed thick brane backgrounds, this convexity condition keeps $e^{A(z)}$ decreasing at large $|z|$ and preserves the convergence of the zero-mode normalization integral,
\begin{equation}
    \int_{-\infty}^{\infty}|\psi_0(z)|^2dz
    =\int_{-\infty}^{\infty}e^{3A(z)}dz<\infty.
\end{equation}
Figure~\ref{figure-bg-zeromode} shows the corresponding zero modes for the unperturbed background and for a symmetric matter perturbation. 
The profile is deformed, but the mode remains localized near the brane. 
This numerical check supports treating the QNM spectral reorganization 
as a change in the massive sector rather than as a loss of localization of the four-dimensional graviton.

\section{Numerical procedures and convergence checks}

\subsection*{Frequency domain.}
Quasinormal modes are defined by the outgoing boundary conditions
\begin{equation}
    \psi(z)\propto 
    \begin{cases}
        e^{i m z}, & z\rightarrow +\infty,\\
        e^{-i m z}, & z\rightarrow -\infty.
    \end{cases}
\end{equation}
The least-damped QNM is denoted by $\bar{m}^{(\epsilon)}$ and is distinct from the graviton zero mode.

For Type I we use the Bernstein spectral method~\cite{Fortuna:2020obg}. 
Following the standard compactification used in thick brane QNM problems, we introduce
\begin{equation}
    u=\frac{\sqrt{k^2 z^2+1}-1}{k z}\in[-1,1],
\end{equation}
so that the infinite $z$ domain is mapped to a compact interval. 
After factoring out the asymptotic outgoing behavior,
\begin{equation}
    \psi(u)=e^{-\frac{i m/k}{u-1}}e^{\frac{i m/k}{u+1}}\tilde{\psi}(u),
\end{equation}
the equation for $\tilde{\psi}(u)$ is regular on $[-1,1]$ and can be solved by expanding it in Bernstein polynomials. 
In practice, we increase the spectral order until the relevant QNMs are stable to the displayed precision.

For Type II we use a shooting method~\cite{Pani:2013pma}. 
Because the potential is even, one may integrate from a large positive cutoff $z_{\max}$ toward the origin with outgoing seed data
\begin{equation}
    \psi(z_{\max})=e^{i m z_{\max}},\qquad \psi'(z_{\max})=i m e^{i m z_{\max}},
\end{equation}
and search the complex $m$ plane for roots satisfying the required parity condition at the origin. 
We retain only roots that remain stable when the integration resolution and $z_{\max}$ are increased.

\begin{figure*}[t]
    \centering
    \includegraphics[width=\textwidth]{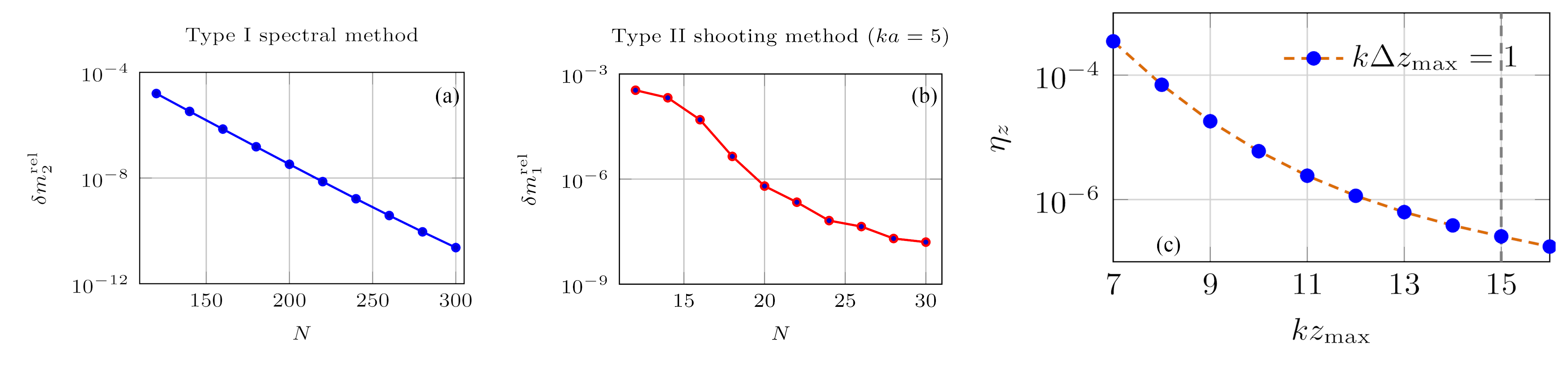}
    \caption{Numerical robustness checks. 
    (a) Relative convergence error defined by Eq.~\eqref{eq:relative-convergence} for the second QNM of Type I perturbations with $w=3$. 
    (b) The same convergence measure for the first (least-damped) Type II QNM obtained by the shooting method with $\epsilon=10^{-3}$, $k=1$, $a=5$, and adjacent shooting resolutions separated by $\Delta N=2$. 
    (c) Cutoff sensitivity of the same Type II QNM, measured by the adjacent relative drift as a function of $kz_{\max}$ with $k\Delta z_{\max}=1$; the gray vertical dashed line marks the adopted working cutoff $kz_{\max}=15$.}
    \label{figure-error}
\end{figure*}

For both methods we monitor the adjacent-resolution drift, where $N$ denotes the numerical resolution parameter (spectral order or shooting resolution). 
Only modes stable under increasing resolution are quoted. 
For parameter choices in which both methods were numerically tractable, representative modes were cross-checked between them.
To compare the spectral and shooting calculations on the same dimensionless scale, Figs.~\ref{figure-error}(a) and~\ref{figure-error}(b) plot the relative convergence measure
\begin{equation}
    \delta m_n^{\mathrm{rel}}
    =\frac{\left|m_n(N+\Delta N)-m_n(N)\right|}{|m_n(N)|}.
    \label{eq:relative-convergence}
\end{equation}
Here $\Delta N$ is the step between neighboring numerical settings. 
For Fig.~\ref{figure-error}(a), $N$ is the Bernstein order and $\Delta N=20$ for the plotted sequence. 
For the Type I calculation shown in Fig.~\ref{figure-error}(a), 
increasing the Bernstein order from $N=120$ to $N=300$ drives the relative error of the second QNM 
from $1.6\times10^{-5}$ to $2.3\times10^{-11}$. 
For the Type II shooting calculation, we parametrize the outer cutoff as
\begin{equation}
    kz_{\max}=ka+c_{\max},
\end{equation}
where $c_{\max}$ is the offset of the outgoing boundary beyond the remote deformation region. 
For the representative Type II run in Figs.~\ref{figure-error}(b) and~\ref{figure-error}(c), 
$\epsilon=10^{-3}$, $k=1$, and $a=5$, so $ka=5$; after the cutoff scan, we adopt $c_{\max}=10$, i.e., $kz_{\max}=15$. 
In the wider scan, the adopted stable offsets for $\epsilon=10^{-5},10^{-4},10^{-3},10^{-2},10^{-1}$ are 
$c_{\max}=5,7,10,15,20$, respectively; empirically, a larger offset is required for convergence as the perturbation amplitude increases.
Figure~\ref{figure-error}(b) shows the shooting-method convergence of the first (least-damped) Type II QNM. 
Figure~\ref{figure-error}(c) then tests the cutoff dependence of the same QNM using
\[
    \begin{aligned}
    \eta_z(x)
    &=\frac{\left|m(x+\Delta x)-m(x)\right|}
    {|m(x)|},\\
    x&=kz_{\max},\qquad \Delta x=k\Delta z_{\max}=1.
    \end{aligned}
\]
Here $m(x)$ is the QNM frequency obtained with the outgoing boundary condition imposed at $x=kz_{\max}$. 
The underlying QNM frequencies are computed at $kz_{\max}=7,8,\ldots,17$, yielding adjacent-drift points at $x=7,8,\ldots,16$, because each $\eta_z(x)$ compares the results at $x$ and $x+1$.
The drift decreases from $3.50\times10^{-4}$ for $kz_{\max}=7\to8$ to $1.75\times10^{-7}$ for $kz_{\max}=16\to17$; at the working cutoff $kz_{\max}=15$ it is already $2.56\times10^{-7}$. 
Thus the displayed Type II QNM is insensitive to further cutoff enlargement, and roots failing these stability checks are discarded.

\subsection*{Time domain.}
For the time-domain evolution we adopt light-cone coordinates $u=t-z$ and $v=t+z$, so that the wave equation becomes
\begin{equation}
    \left[4 \frac{\partial^2}{\partial_u \partial_v}+V(u,v)\right]\Psi(u,v)=0.
\end{equation}
The evolution is performed on a uniform null grid by a standard characteristic finite-difference scheme. 
For a grid spacing $h$, the value at the future point $N=(u+h,v+h)$ 
is updated from the other three corners of the corresponding null-grid cell, yielding a second-order accurate evolution of the waveform.

We initialize the evolution with odd Gaussian data,
\begin{align}
    \Psi_{\text{in}}(u,0)&=\sin(ku)e^{-k^2u^2/2},\\
    \Psi_{\text{in}}(0,v)&=\sin(-kv)e^{-k^2v^2/2}.
\end{align}
This choice is deliberate: for a symmetric potential the graviton zero mode is even, so odd initial data suppress the protected zero mode and make the QNM response easier to isolate. 
It therefore provides a clean diagnostic of spectral migration before source-dependent astrophysical effects are included.
If the initial data contain an even component, the zero mode and continuum tail are excited together with the QNMs. 
Such data are physically allowed, but they obscure the diagnostic contrast between the prompt ringdown and the delayed echo response. 
The waveform statements in the article should therefore be understood as statements about 
a QNM-dominated diagnostic channel, rather than as a claim that every possible source profile isolates the same modes with the same visibility.

Waveforms are extracted at fixed $z=z_{\text{ext}}$ in the asymptotic region. 
The values used in the main text, $kz_{\text{ext}}=5$ for Type I and $kz_{\text{ext}}=20$ for Type II, are chosen to lie outside the main scattering barrier. 
We have verified that varying the extraction point within the asymptotic region changes the overall arrival time and amplitude, but not the qualitative separation between prompt ringdown and late echoes.

The ringdown frequencies are extracted by fitting the waveform with
\begin{equation}
    \Psi(t)=\sum_{\ell=1}^{K} A_\ell e^{\operatorname{Im}(\omega_\ell)t}
    \sin\left[\operatorname{Re}(\omega_\ell)t-\theta_\ell\right].
\end{equation}
Here $K$ is the number of fitted damped components, while $A_\ell$ and $\theta_\ell$ are their amplitudes and phases, respectively.
Throughout the article, we set $p=0$, so that $m_\ell=\omega_\ell$. 
For prompt-ringdown fits we use $K=1$ as the default and repeat selected fits with $K=2$ as a stability check. 
For the delayed Type II echo windows we instead use multi-component fits with $K=3$ or $K=4$, because the original prompt remnant, the cavity-induced long-lived component, and tail/overlap contributions can coexist in the same window. 
The multiple time-domain points shown in the Type II frequency-domain/time-domain comparison in the main text are therefore individual fitted components from these echo-window fits, not the result of a single $K=1$ fit.
The time origin $\Delta t=0$ is set at the peak of the initially arriving burst. 
For Type I, the default window is $10<k\Delta t<20$; moving either endpoint within the clean exponentially damped part of the signal changes only the last quoted digits of the Type I values reported in the main article. 
For Type II, the prompt fit is performed after the burst but before the first round-trip echo, that is, within $k\Delta t<2ka$. 
The delayed echoes are then fitted in separate windows centered near $\Delta t\simeq2a,4a,\ldots$. 
This separation is necessary because the original prompt branch, the new long-lived cavity branch, 
and the tail can overlap at late times. 
The first Type II echo is of order $\epsilon$ relative to the prompt response in the controlled 
scattering problem, with later echoes being further suppressed; for $\epsilon=10^{-3}$--$10^{-2}$ the cleanest spectral-instability signal is therefore expected in a low-amplitude part of the waveform.

Time-domain fitting in this problem is therefore used to identify which spectral features are manifested in the waveform and at what stage of the evolution they appear. 
For Type I, the prompt ringdown follows the least-damped branch even though higher overtones move nonperturbatively. 
For Type II, the prompt ringdown is resilient, while the echo train carries the imprint of the new long-lived branch generated by the remote deformation.

\end{document}